\documentclass[aps,prl,showpacs,preprint,amsmath,amssymb,superscriptaddress]{revtex4}

\usepackage{graphicx}
\begin{document}
\bibliographystyle{prsty}
\title{Local electronic structure of Cr in the II-VI diluted ferromagnetic semiconductor Zn$_{1-x}$Cr$_{x}$Te}

\author{M.~Kobayashi}
\altaffiliation{Present address: Department of Applied Chemistry, 
School of Engineering, University of Tokyo, 7-3-1 Hongo, Bunkyo-ku, 
Tokyo 113-8656, Japan}
\affiliation{Department of Physics, University of Tokyo, 
7-3-1 Hongo, Bunkyo-ku, Tokyo 113-0033, Japan}
\author{Y.~Ishida}
\altaffiliation{Present address: RIKEN, SPring-8 Center, Sayo-cho, Hyogo 679-5148, Japan}
\affiliation{Department of Physics, University of Tokyo, 
7-3-1 Hongo, Bunkyo-ku, Tokyo 113-0033, Japan}
\author{J.~I.~Hwang}
\affiliation{Department of Physics, University of Tokyo, 
7-3-1 Hongo, Bunkyo-ku, Tokyo 113-0033, Japan}
\author{G.~S.~Song}
\affiliation{Department of Physics, University of Tokyo, 
7-3-1 Hongo, Bunkyo-ku, Tokyo 113-0033, Japan}
\author{A.~Fujimori}
\affiliation{Department of Physics, University of Tokyo, 
7-3-1 Hongo, Bunkyo-ku, Tokyo 113-0033, Japan}
\author{C.-S.~Yang}
\affiliation{National Synchrotron Radiation Research Center, 
Hsinchu 30076, Taiwan}
\author{L.~Lee}
\affiliation{National Synchrotron Radiation Research Center, 
Hsinchu 30076, Taiwan}
\author{H.-J.~Lin}
\affiliation{National Synchrotron Radiation Research Center, 
Hsinchu 30076, Taiwan}
\author{D.~J.~Huang}
\affiliation{National Synchrotron Radiation Research Center, 
Hsinchu 30076, Taiwan}
\author{C.~T.~Chen}
\affiliation{National Synchrotron Radiation Research Center, 
Hsinchu 30076, Taiwan}
\author{Y.~Takeda}
\affiliation{Synchrotron Radiation Research Unit, 
Japan Atomic Energy Agency, Sayo-gun, Hyogo 679-5148, Japan}
\author{K.~Terai}
\affiliation{Synchrotron Radiation Research Unit, 
Japan Atomic Energy Agency, Sayo-gun, Hyogo 679-5148, Japan}
\author{S.-I.~Fujimori}
\affiliation{Synchrotron Radiation Research Unit, 
Japan Atomic Energy Agency, Sayo-gun, Hyogo 679-5148, Japan}
\author{T.~Okane}
\affiliation{Synchrotron Radiation Research Unit, 
Japan Atomic Energy Agency, Sayo-gun, Hyogo 679-5148, Japan}
\author{Y.~Saitoh}
\affiliation{Synchrotron Radiation Research Unit, 
Japan Atomic Energy Agency, Sayo-gun, Hyogo 679-5148, Japan}
\author{H.~Yamagami}
\affiliation{Synchrotron Radiation Research Unit, 
Japan Atomic Energy Agency, Sayo-gun, Hyogo 679-5148, Japan}
\author{K.~Kobayashi}
\affiliation{Japan Synchrotron Radiation Research Institute, 1-1-1 Kouto, 
Mikaduki-cho, Sayou-gun, Hyogo 679-5148, Japan}
\author{A.~Tanaka}
\affiliation{Department of Quantum Matter, ADSM, 
Hiroshima University, Hagashi-Hiroshima 739-8530, Japan}
\author{H.~Saito}
\affiliation{Nanoelectronics Research Institute, AIST, Tsukuba Central 2, 
Umezono 1-1-1, Tsukuba Ibaraki 305-8568, Japan}
\author{K.~Ando}
\affiliation{Nanoelectronics Research Institute, AIST, Tsukuba Central 2, 
Umezono 1-1-1, Tsukuba Ibaraki 305-8568, Japan}
\date{\today}

\begin{abstract}
The electronic structure of the Cr ions in the diluted ferromagnetic semiconductor Zn$_{1-x}$Cr$_x$Te ($x=0.03$ and 0.15) thin films has been investigated using x-ray magnetic circular dichroism (XMCD) and photoemission spectroscopy (PES). Magnetic-field ($H$) and temperature ($T$) dependences of the Cr $2p$ XMCD spectra well correspond to the magnetization measured by a SQUID magnetometer. The line shape of the Cr $2p$ XMCD spectra is independent of $H$, $T$, and $x$, indicating that the ferromagnetism is originated from the same electronic states of the Cr ion. Cluster-model analysis indicates that although there are two or more kinds of Cr ions in the Zn$_{1-x}$Cr$_x$Te samples, the ferromagnetic XMCD signal is originated from Cr ions substituted for the Zn site. The Cr $3d$ partial density of states extracted using Cr $2p \to 3d$ resonant PES shows a broad feature near the top of the valence band, suggesting strong $s$,$p$-$d$ hybridization. No density of states is detected at the Fermi level, consistent with their insulating behavior. Based on these findings, we conclude that double exchange mechanism cannot explain the ferromagnetism in Zn$_{1-x}$Cr$_{x}$Te. 
\end{abstract}

\pacs{75.50.Pp, 75.30.Hx, 78.70.Dm, 79.60.-i}

\maketitle

\section{Introduction}
Ferromagnetic diluted magnetic semiconductors (DMS's) have opened a way for the manipulation of the spin degree of freedom of electrons through interaction between the local moments of doped magnetic ions and the spins of charge carriers in the host semiconductors. Therefore, ferromagnetic DMS's have been considered to be key materials for semiconductor spin electronics or spintronics \cite{Science_98_Ohno, JSAPI_02_Ohno}, which is intended to manipulate both the charge and spin degrees of freedom of electrons in semiconductors. If ferromagnetism occurs as a result of interaction between the local magnetic moments of doped ions and the spins of charge carriers, the magnetism is called carrier-induced ferromagnetism, and the III-V DMS's Ga$_{1-x}$Mn$_x$As and In$_{1-x}$Mn$_x$As are prototypical systems of carrier-induced ferromagnetism \cite{RMP_06_Jungwirth}. Using the III-V DMS's, new functional devices such as circular polarized light detectors \cite{PRL_78_Koshihara}, spin-related light-emitting diodes \cite{Nature_402_Fiederling}, and field-effect transistors controlling ferromagnetism \cite{Nature_00_Ohno} have been fabricated. However, these devices only act at low temperatures since the Curie temperatures ($T_\mathrm{C}$'s) of the DMS's are below room temperature. Therefore, ferromagnetic DMS's having $T_\mathrm{C}$ above room temperature are strongly desired for practical applications of spintronic devices. 
Ever since the theoretical prediction of ferromagnetism having $T_\mathrm{C}$ exceeding room temperature in wide-gap semiconductor-based DMS's \cite{Science_00_Dietl}, there have been many reports on room-temperature ferromagnetism of wide-gap DMS's such as Ga$_{1-x}$Mn$_x$N and Zn$_{1-x}$Co$_x$O \cite{JMSME_05_Liu}. 
In order to see whether the ferromagnetic properties are intrinsic or extrinsic, anomalous Hall effects \cite{NatMater_04_Toyosaki}, magnetic circular dichroism (MCD) in visible-to-ultraviolet region \cite{Science_06_Ando}, and carrier-doping dependence of the ferromagnetism \cite{AdvMater_04_Schwartz, PRL_05_Kittilstved} have been studied since these properties of DMS are derived from interaction between the host semiconductor and the doped magnetic ions.

\begin{figure}[t!]
\begin{center}
\includegraphics[width=13cm]{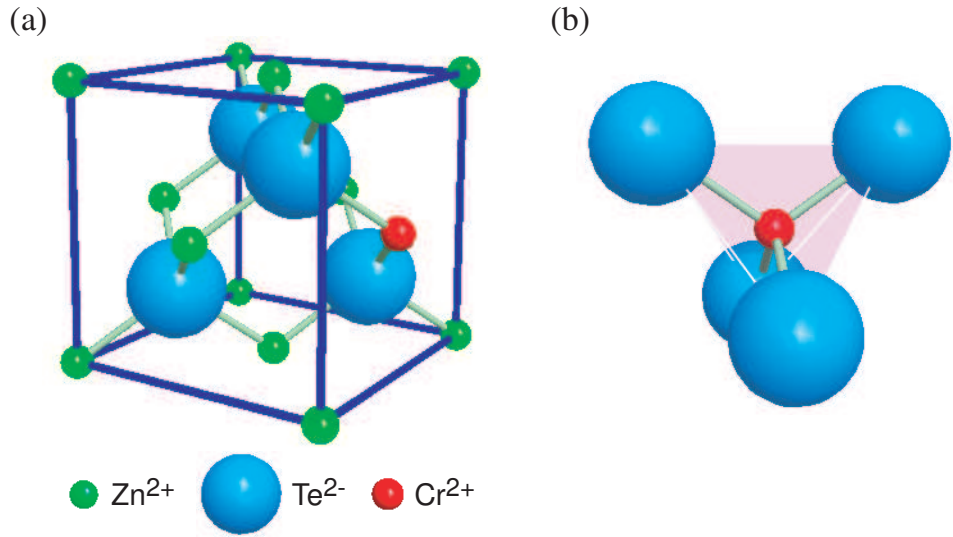}
\caption{Crystal structure of Zn$_{1-x}$Cr$_x$Te. 
(a) Zinc-blend structure. 
(b) [CrTe$_4$]$^{6-}$ cluster. 
}
\label{ZnTeCrystal}
\end{center}
\end{figure}

The II-VI semiconductor ZnTe crystallizes in the zinc-blend structure as shown in Fig.~\ref{ZnTeCrystal}(a), has a band gap of $\sim 2.4$ eV, and shows $p$-type electrical conductivity. Cr-doped ZnTe crystal, in which the Cr concentration is below 1\%, has been investigated before the discovery of ferromagnetism in heavily Cr-doped ZnTe thin films. 
Infrared absorption \cite{PRB_70_Vallin} and electron spin resonance \cite{PRB_74_Vallin} studies of bulk Zn$_{1-x}$Cr$_x$Te have suggested that the Cr ions are divalent and are subjected to tetragonal Jahn-Teller distortion. 
MCD measurements in visible-to-ultraviolet region on bulk Zn$_{1-x}$Cr$_x$Te crystals have revealed a positive $p$-$d$ exchange constant $N\beta$, that is, the exchange interaction between the hole spin and the local magnetic moment is ferromagnetic \cite{PRB_96_Mac}. 
Recently, Saito {\it et al}. \cite{PRL_03_Saito} have succeeded to prepare ZnTe thin films doped with high concentration of Cr atoms ($x \sim 20$\%) by the molecular beam epitaxy (MBE) technique. 
The Zn$_{1-x}$Cr$_x$Te thin films showed ferromagnetism at room temperature and their MCD signals observed at the absorption edge of ZnTe showed magnetic-field ($H$) and temperature ($T$) dependences which follow these of magnetization ($M$), indicating that there is strong interaction between the spins of host $s$,$p$-band electrons and the magnetic moments of the doped Cr ions \cite{PRL_03_Saito, JAP_03_Saito}. Therefore, Zn$_{1-x}$Cr$_x$Te has attracted much attention as an intrinsic DMS with strong $s$,$p$-$d$ interaction.

The ferromagnetic properties of Zn$_{1-x}$Cr$_{x}$Te thin films have been investigated so far \cite{STAM_05_Kuroda}. While thin films with high Cr concentrations have shown a clear hysteresis loop, those with low Cr concentrations have exhibited unusual hysteresis: the loop of the hysteresis becomes thin near the zero field in the $M$-$H$ curve, which can be explained by a superposition of ferromagnetic and superparamagnetic components. 
In the $M$-$T$ curves of ferromagnetic Zn$_{1-x}$Cr$_{x}$Te thin films, blocking phenomena, namely, difference in the magnetization between the zero-field-cool (ZFC) and field-cool (FC) have also been observed. The blocking temperature ($T_B$), where the ZFC and FC curves start to separate, increased with increasing Cr concentration: $T_B \sim 5$ K for $x=0.01$ and $\sim 95$ K for $x=0.17$. 
These results imply that ferromagnetic Zn$_{1-x}$Cr$_x$Te is a magnetically random system with some antiferromagnetic interaction between the Cr ions and that the magnetism is sensitive to the Cr concentration.

Effects of doping on the ferromagnetism of Zn$_{1-x}$Cr$_x$Te have been studied \cite{PRL_06_Ozaki, APL_05_Ozaki, NatMater_07_Kuroda}. Iodine (I), which is expected to be an electron dopant, enhances the ferromagnetism while nitrogen (N), which is expected to be a hole dopant, suppresses it \cite{PRL_06_Ozaki, APL_05_Ozaki}. These effects have been explained based on the double-exchange mechanism \cite{PRL_06_Ozaki, JJAP_01c_Sato}. 
However, carrier-induced ferromagnetism in Zn$_{1-x}$Cr$_x$Te is doubted because the Zn$_{1-x}$Cr$_x$Te films are highly insulating. 
Furthermore, the tendency that N doping increases hole carrier concentration and suppresses the ferromagnetism is opposite to the observation for Ga$_{1-x}$Mn$_{x}$As \cite{PRB_98_Matsukura}, in which the ferromagnetic property is enhanced by the increase of hole concentration. 
Recently, spatially inhomogeneous distributions of the Cr ions have been pointed out to influence the magnetic properties \cite{JPCM_04_Ozaki, JJAP_06_Fukushima, AIP_07_Ofuchi}. 
Spatially resolved energy-dispersive x-ray spectroscopy study has recently revealed that co-doping with I induces inhomogeneous formation of Cr-rich (Zn,Cr)Te nano-regions, whereas co-doping with N results in homogeneous Cr-ion distributions \cite{NatMater_07_Kuroda}.


X-ray magnetic circular dichroism (XMCD) and resonant photoemission spectroscopy (RPES) are powerful tools to investigate the electronic structure of DMS \cite{PRB_99_Okabayashi, PRB_02_Mizokawa, PRL_03_Keavney, JESRP_04_Mizokawa, PRB_05_Edmond, JESRP_05_Rader, JESRP_05_Fujimori, PRB_05_Hwang, PRB_05_Kobayashi, PRB_06_Mamiya}. XMCD is defined as the difference between the core-level x-ray absorption spectroscopy (XAS) spectra taken with right-handled ($\mu^+$) and left-handled ($\mu^-$) circularly polarized x rays. 
Because XMCD is sensitive only to magnetically active species, it is very efficient to extract information about the electronic and magnetic properties of doped magnetic ions. 
In RPES, when the incident photon energy is adjusted to the $2p \to 3d$ core excitation energy, the photoemission intensity of the $3d$ partial density of states (PDOS) is resonantly enhanced \cite{PRB_04_Rader, PRL_91_Tjeng}. 

Our previous Cr $2p$ XMCD measurements on Zn$_{1-x}$Cr$_{x}$Te ($x=0.045$) thin film \cite{APEX_08_Ishida} have revealed that the orbital moment of the Cr $3d$ electrons is largely quenched compared with the value of the Cr$^{2+}$ ion in the tetrahedral crystal field. 
The XMCD intensity increased with increasing $H$ up to 7 T, indicating the existence of paramagnetism and/or superparamagnetism in Zn$_{1-x}$Cr$_x$Te, and that the magnetically active Cr ions had a single chemical environment although a small amount of magnetically inactive Cr ions existed. Atomic multiplet theory analysis has suggested that the Cr ions are divalent and are subjected to tetragonal Jahn-Teller distortion, whose distortion axes are equally distributed in the $X$, $Y$ and $Z$ directions. The valence-band PES spectra showed suppressed spectral weight near the Fermi level ($E_\mathrm{F}$) \cite{APEX_08_Ishida}. Based on these observations, we have proposed that the spectral suppression is originated from the Jahn-Teller distortion and/or Coulomb interaction between Cr $3d$ electrons.

However, some points remain to be confirmed concerning the above suggestions. Since the XMCD spectra were recorded only in the applied magnetic field parallel to the $c$ axis, the direction of Jahn-Teller distortion was not confirmed experimentally. 
In addition, the effects of inhomogeneous Cr distribution in Zn$_{1-x}$Cr$_{x}$Te on the electronic structure of the Cr ions should be elucidated. 
In order to address these issues, we have performed $x$, $H$, $T$, and incident-photon angular dependences of XAS and XMCD on ferromagnetic Zn$_{1-x}$Cr$_x$Te. In this paper, we repot on the results of XMCD and PES measurements combined with CI cluster-model analysis on Zn$_{1-x}$Cr$_{x}$Te thin films having different Cr concentrations. 
The $x$ and $H$ dependences of XMCD spectra compared with $M$ measured by a SQUID magnetometer provided experimental evidence that we measured the bulk magnetic properties. The $T$ and angle dependences of XMCD, the former of which is sensitive to the orbital degree of freedom and the latter of which is sensitive to the direction of distortion, revealed quenching of the orbital moment and isotropic electronic configuration of the Cr ions. In addition, the CI cluster-model analysis suggested that the Cr ions are subjected to Jahn-Teller distortion with isotropic distribution of the distortion axes and the ferromagnetism is caused by the Cr ions substituted for the Zn sites. The $p$-$d$ exchange coupling constant $N\beta$ estimated from the electronic structure parameters was consistent with the MCD measurements. The Cr $3d$ PDOS obtained for the $2p \to 3d$ RPES was suppressed in the whole Cr concentrations, implying that carrier-induced ferromagnetism is not effective in Zn$_{1-x}$Cr$_x$Te. 

\section{Experimental}
Zn$_{1-x}$Cr$_x$Te ($x=0.03$ and 0.15) thin films were grown on insulating GaAs(001) substrates by the MBE technique. The total thickness of the Zn$_{1-x}$Cr$_x$Te thin films was 50 nm on a 20 nm ZnTe buffer layer. During the deposition, the substrate was kept at a temperature of $\sim 250$ $^{\circ}$C. Details of the sample preparation were described in \cite{PRL_03_Saito}. Since XMCD is a surface sensitive probe, the films were covered by an Al capping layer of 2 nm thickness so as to avoid surface oxidization. Ferromagnetism was confirmed by magnetization measurements using a SQUID magnetometer (Quantum Design, Co. Ltd). The Curie temperatures of the $x=0.03$ and 0.15 films were estimated to be $\sim 150$ and $\sim 280$ K, respectively.

XMCD measurements were performed at the Dragon Beamline BL11A of National Synchrotron Radiation Research Center in the total-electron-yield mode. An XMCD hysteresis loop at the Cr $2p_{3/2}$ edge was measured by the total-fluorescence-yield mode. The monochromator resolution was $E/{\Delta}E \textgreater 10\,000$. The polarization of incident photons was fixed and XMCD ($\mu^{+} - \mu^{-}$) spectra were obtained by switching the magnetic field.  The background of the XAS spectra at the Cr $2p$ edge was assumed to be a superposition of the spectrum of ZnTe in this region and  hyperbolic tangent functions. 
The Cr $2p \to 3d$ RPES measurements were performed at BL23SU of SPring-8 using a Gammadata Scienta SES-2002 hemispherical analyzer operated in the transmission mode. The monochromator resolution was $E/{\Delta}E \textgreater 10\,000$. The spectra were taken at room temperature in a vacuum below $5.0 \times 10^{-8}$ Pa. The total resolution of the spectrometer including temperature broadening was $\sim 200$ meV.

\section{Soft x-ray magnetic circular dichroism}

As described in the introduction, the ferromagnetism of Zn$_{1-x}$Cr$_{x}$Te is enhanced with Cr concentration. From the magnetization measurements described below, the enhancement may be related to the inhomogeneous Cr distribution. A question then arises what happens in the electronic structure related to the ferromagnetism if a high concentration of Cr ions are doped into a ZnTe thin film. The $x$ dependence of XMCD measurement can approach this problem. In addition, we have measured $H$, $T$, and angle dependences of XMCD and XAS so as to obtain the information of the electronic structure of the Cr ions.

\subsection{Cr-concentration dependence}
\begin{figure}[t!]
\begin{center}
\includegraphics[width=13.5cm]{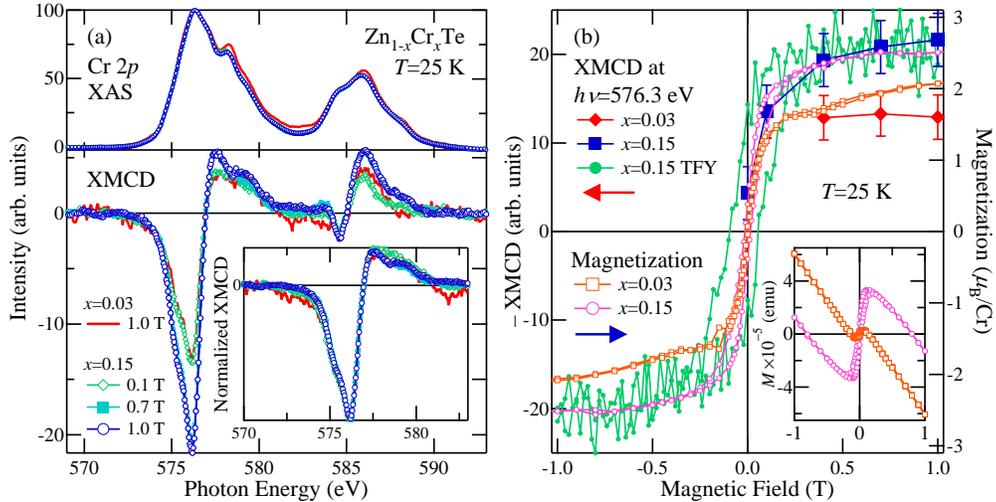}
\caption{Magnetic-field dependence of the Cr $2p$ XMCD spectra of Zn$_{1-x}$Cr$_x$Te thin films. 
(a) XMCD spectra at various $H$. 
(b) XMCD intensity as a function of $H$ compared with the magnetization curves. The diamagnetic component of the substrate has been subtracted from the raw magnetization curves (inset). The XMCD hysteresis loop has been measured by the total-fluorescence-yield (TFY) mode. 
}
\label{XMCD_H-dep}
\end{center}
\end{figure}

In order to investigate the Cr-concentration dependence of the ferromagnetism in Zn$_{1-x}$Cr$_{x}$Te, we compare in Fig.~\ref{XMCD_H-dep}(a) the Cr $2p$ XMCD spectra of the Zn$_{1-x}$Cr$_{x}$Te thin films of $x=0.03$ and 0.15 recorded under a fixed condition, $H=1.0$ T and $T=25$ K. 
Here, the spectra of different $x$'s have been normalized to the XAS [$(\mu^+ + \mu^-)/2$] peak height at $\sim 576$ eV. 
Both the XAS and XMCD spectra of the $x=0.03$ and 0.15 samples have nearly identical line shapes except in the region around 578 eV, where the XAS spectrum of the $x=0.03$ film is slightly stronger than that of the $x=0.15$ one, and at $H=1.0$ T the XMCD intensity increases with increasing $x$. Both the XAS and XMCD spectra are similar to those of a $x=0.45$ sample previously reported \cite{APEX_08_Ishida}. 
These results suggest that although there may be two or more kinds of Cr ions having different electronic structures as observed by XAS, the ferromagnetic properties of these films reflected on the XMCD spectra are originated from a common magnetically active component. One can also see in Fig.~\ref{XMCD_H-dep}(a) that the magnetically active component of the $x=0.15$ film shows a stronger XMCD intensity than that of the $x=0.03$ film. 
Considering that XMCD is sensitive only to local electronic states, the increase of the magnetization per Cr ion with Cr concentration suggests that the distance between the Cr ions affects the magnetism of (Zn,Cr)Te.

\subsection{Magnetic-field dependence}
XMCD measurements with varying $H$ are useful to investigate chemically and magnetically inhomogeneous samples \cite{JESRP_05_Rader, JESRP_05_Fujimori}. At low $H$, the XMCD spectra predominantly reflect the ferromagnetic component of the Cr ions while at high $H$, the paramagnetic component is superimposed. Figure~\ref{XMCD_H-dep}(a) shows the $H$ dependence of the XMCD spectra of the $x=0.15$ film measured at $T=25$ K. 
The XMCD intensity increases with increasing $H$, while the spectra maintain the same line shape up to $H=1.0$ T, as shown in the inset of Fig.~\ref{XMCD_H-dep}(a), where the XMCD spectra have been normalized to the negative peak at 576 eV. In magnetic fields which are sufficient to saturate the magnetization of the ferromagnetic component, the XMCD intensity hardly increases with increasing $H$, indicating that there are little paramagnetic magnetization in both the $x=0.03$ and $x=0.15$ thin films below $H=1.0$ T, consistent with the previous XMCD measurements \cite{APEX_08_Ishida}. 
These results suggest that the XMCD signal is dominated by the ferromagnetic one below $H=1.0$ T, although at high fields the paramagnetic (or superparamagnetic) and antiferromagnetic XMCD signals become detectable \cite{APEX_08_Ishida}. 
In order to see the consistency between the macroscopic and microscopic magnetic measurements, comparison of the XMCD intensity with the magnetization $M$ is made in Fig.~\ref{XMCD_H-dep}(b). The ratio of the XMCD intensity to $M$ at high magnetic fields almost coincide between the different films, providing experimental evidence that the XMCD intensity reflects the bulk magnetic properties of Zn$_{1-x}$Cr$_x$Te. 
The good agreement between the $M$-$H$ curve and the XMCD intensity as a function of $H$ gives evidence for ferromagnetism induced by the Cr ion. 
It is therefore likely that majority of the doped Cr ions in Zn$_{1-x}$Cr$_x$Te magnetically interact with each other and give rise to the ferromagnetic behavior.

\subsection{Temperature dependence}
\begin{figure}[t!]
\begin{center}
\includegraphics[width=8.0cm]{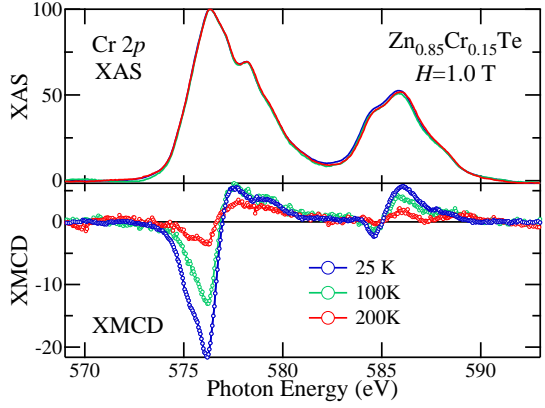}
\caption{Temperature dependence of the Cr $2p$ XMCD spectra of the Zn$_{1-x}$Cr$_x$Te ($x=0.15$) thin film at $H=1.0$ T. 
}
\label{TempDep}
\end{center}
\end{figure}

For a $3d$ transition-metal ion in a tetrahedral-symmetry ($T_d$) crystal field with an open $t_2$ shell, a strong temperature dependence of the XAS spectra is expected due to the degeneracy of the orbital degree of freedom in the ground state. For the Cr$^{2+}$ ion having the $d^4$ electronic configuration in the $T_d$ crystal field, the $t_2$ states are partially occupied and the orbital degree of freedom survives. However, the Cr $L_{2,3}$ XAS spectra of Zn$_{1-x}$Cr$_x$Te did not change with temperature as shown in Fig.~\ref{TempDep}. This result implies that the orbital degeneracy is lifted due to a Jahn-Teller distortion which splits the $t_2$ levels \cite{PRL_05_Haverkort}. Alternatively, the $3d$ shell may be completely filled, i.e., the Cr ion is in the Cr$^+$ state and has the $d^5$ electronic configuration.

\subsection{XMCD sum rules}
\begin{figure}[t!]
\begin{center}
\includegraphics[width=13.5cm]{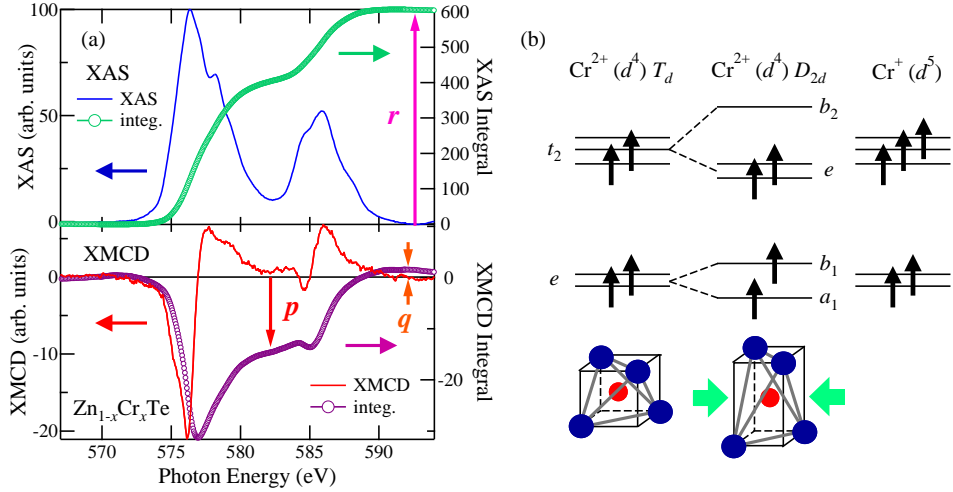}
\caption{Application of the XMCD sum rules to the Cr $2p$ XAS and XMCD spectra of Zn$_{1-x}$Cr$_{x}$Te thin films. 
(a) XAS and XMCD spectra and their spectral integrals. 
(b) Candidates for the electronic structure of the Cr ions. 
The orbital degree of freedom of the Cr$^{2+}$ ion in the $T_d$ crystal field (left column) is quenched in the Cr$^{2+}$ ion in the $D_{2d}$ crystal field (middle column) or in the Cr$^{+}$ configuration (right column). 
}
\label{XMCD_SumRules}
\end{center}
\end{figure}

By applying the XMCD sum rules \cite{PRL_93_Carra, PRL_92_Thole}, one can estimate the spin ($M_\mathrm{spin}$) and orbital magnetic moments ($M_\mathrm{orb}$) of the Cr ion separately by
\begin{equation}
M_\mathrm{orb} = -\frac{2q}{3r} \left( 10 - N_d \right), 
\end{equation}
\begin{equation}
M_\mathrm{spin} + 7M_\mathrm{T} = -\frac{3p - 2q}{r} \left( 10 - N_d \right),
\end{equation}
where $p$, $q$, and $r$ are the integrated intensities of the XAS and XMCD spectra as shown in Fig.~\ref{XMCD_SumRules}(a), $N_d$ is the number of $3d$ electrons, and $M_\mathrm{T}$ is the expectation value of the magnetic dipole operator, which is negligibly small with respect to $M_\mathrm{spin}$ because of relatively weak spin-orbital coupling in $3d$ electrons \cite{PRL_93_Carra}. 
The ratio $M_\mathrm{orb}/ M_\mathrm{spin}$ was estimated to be $\sim 0.11 \pm 0.12$, indicating that the orbital moment is significantly suppressed, consistent with the previous XMCD measurements \cite{APEX_08_Ishida}. 
Thus, the candidate electronic structures for the magnetically active Cr ions are the Cr$^{2+}$ ion in $D_{2d}$ symmetry or Cr$^{+}$ ($d^5$) configuration, as shown in Fig.~\ref{XMCD_SumRules}(b). Below, we shall discuss the electronic structure of the Cr ions under these constraints. 

\subsection{Incident photon angle dependence}
\begin{figure}[t!]
\begin{center}
\includegraphics[width=13.5cm]{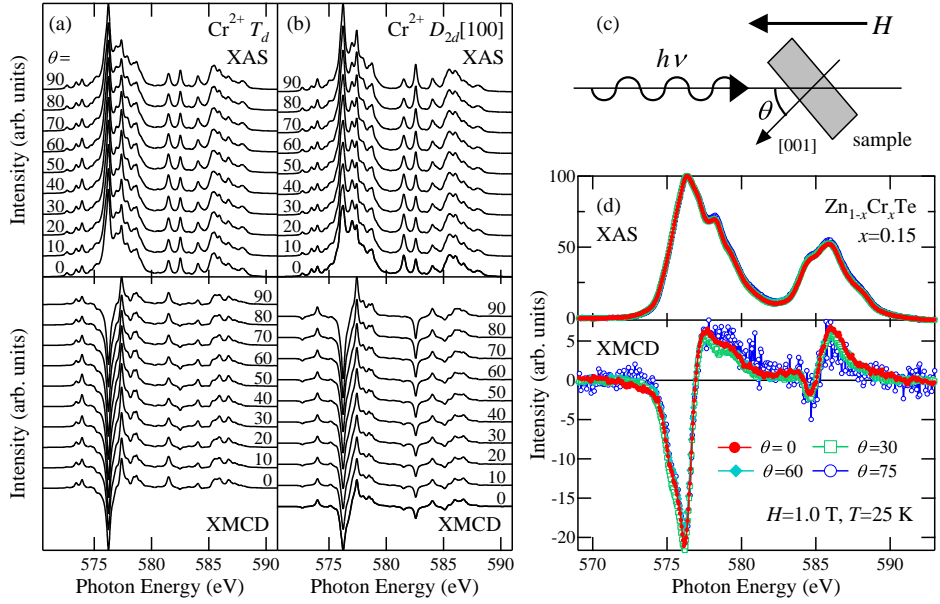}
\caption{Incident photon angle dependence of the Cr $2p$ XAS and XMCD spectra of Zn$_{1-x}$Cr$_{x}$Te thin films. 
(a), (b) XAS and XMCD spectra calculated using atomic multiplet theory for Cr$^{2+}$ $T_d$ and $D_{2d}$ with varying $\theta$. (c) Experimental setup for the angle-dependent measurements. (d) Cr $2p$ XAS and XMCD spectra with various $\theta$. 
}
\label{AngleDep}
\end{center}
\end{figure}

When the Cr$^{2+}$ ion is under the uniaxial distortion of tetragonal symmetry $D_{2d}$, the electronic state should have an anisotropy. Calculations using atomic multiplet theory for the $T_d$ and $D_{2d}$ symmetries have been performed by varying the incident angle $\theta$ of x ray as shown in Fig.s~\ref{AngleDep}(a) and (b). Here, $\theta=0^\circ$ is defined by the angle between the surface normal and the propagation vector of the incident x rays as shown in Fig.~\ref{AngleDep}(c). Racah parameters have been assumed to be the same as those of the Cr$^{2+}$ ion. 
Both the XAS and XMCD spectra for $T_d$ are independent of $\theta$, while the spectra for $D_{2d}$ show systematic changes with $\theta$. The analysis thus indicates that if the Cr ion is under the uniaxial distortion, the XAS and XMCD spectra would show $\theta$ dependence. 

Figure~\ref{AngleDep}(d) shows the XAS and XMCD spectra measured at several $\theta$'s. 
There is little $\theta$ dependence in the XAS and XMCD spectra of the $x=0.15$ film, suggesting that the ferromagnetic Cr ion consists of an almost isotropic electronic configuration. 
Taking into account the quenching of the orbital moment, the observation implies an equal distribution of the axis of the Jahn-Teller distortion in the $a$-, $b$-, and $c$-directions, or the $d^5$ electronic configuration of the Cr ion.

\section{Configuration-interaction cluster-model calculation}
The substitutional magnetic ion in a II-VI DMS is tetrahedrally coordinated by anions as shown in Fig.~\ref{ZnTeCrystal}(b), and therefore the electronic structure of the magnetic ion is influenced by the crystal field of the ligand ions and hybridization with ligand orbitals. In order to determine the electronic structure of the ferromagnetic component, we have performed CI cluster-model analysis for the Cr $2p$ XAS and XMCD spectra. 
While atomic multiplet theory treats the symmetry and strength of the crystal field through the strength of the crystal-field splitting of the energy levels of one-electron orbitals, hybridization between ligand and $3d$ orbitals is explicitly taken into account in CI cluster model. Here, CI means interaction between different charge-transfer electronic configurations. Therefore, by applying CI cluster-model analysis, one can obtain more detailed information of electronic structure of the Cr ions such as $N\beta$. 
CI cluster-model analysis is useful for describing the local electronic structure of $3d$ ions in DMS and enables us to estimate their electronic structure parameters: charge-transfer energy $\Delta$, $d$-$d$ Coulomb interaction energy $U_{dd}$, and Slater-Koster parameter ($pd\sigma$).

Let us first consider the possibility of the $d^{4}$ electronic configuration. According to reports on the dependence of the lattice constant on the Cr concentration in Zn$_{1-x}$Cr$_{x}$Te \cite{STAM_05_Kuroda, PRB_02_Saito}, at low Cr concentrations ($x \lesssim 0.04$), the $x$ dependence of the lattice constant obeys Vegard's law, indicating that the Cr ions substitute for the cation sites. At higher Cr concentrations ($x \gtrsim 0.05$), the lattice constant of Zn$_{1-x}$Cr$_{x}$Te becomes the same as that of ZnTe and Zn$_{1-x}$Cr$_{x}$Te shows cubic symmetry due to relaxation of the lattice constant. Therefore, if the Cr$^{2+}$ ions are accommodated in the Zn$_{1-x}$Cr$_{x}$Te lattice, it is likely that the Jahn-Teller axes along [001], [010], and [100] are distributed isotropically as a consequence of the relaxation of lattice constant.

Next, let us examine the possibility of the $d^{5}$ electronic configuration. Recently, a Raman electron paramagnetic resonance study has reported that Cr$^{+}$ ions existed in bulk Zn$_{1-x}$Cr$_{x}$Te crystal ($x < 0.02$) as an acceptor although Cr$^{+}$ is a minority electronic configuration in contrast to the dominant Cr$^{2+}$ one \cite{PRB_07_Lu}. 
According to a previous MCD measurement on Zn$_{1-x}$Cr$_{x}$Te thin film \cite{Magneto-Optics}, the $p$-$d$ exchange constant $N\beta$ is positive (ferromagnetic) like bulk Zn$_{1-x}$Cr$_{x}$Te. For the half-filled or more than half-filled $3d$ orbitals as realized in Ga$_{1-x}$Mn$_{x}$As \cite{JAP_98_Ando} and Zn$_{1-x}$Co$_{x}$Te \cite{Magneto-Optics}, the $p$-$d$ exchange becomes negative (antiferromagnetic) because the up-spin states of the $3d$ orbitals are fully occupied so that only a down-spin $p$ electron can hop into the $3d$ orbitals. 
It follows from these reports that the half-filled electronic configuration Cr$^{+}$ is minority in Zn$_{1-x}$Cr$_{x}$Te, if it exists.

\begin{figure}[t!]
\begin{center}
\includegraphics[width=13.5cm]{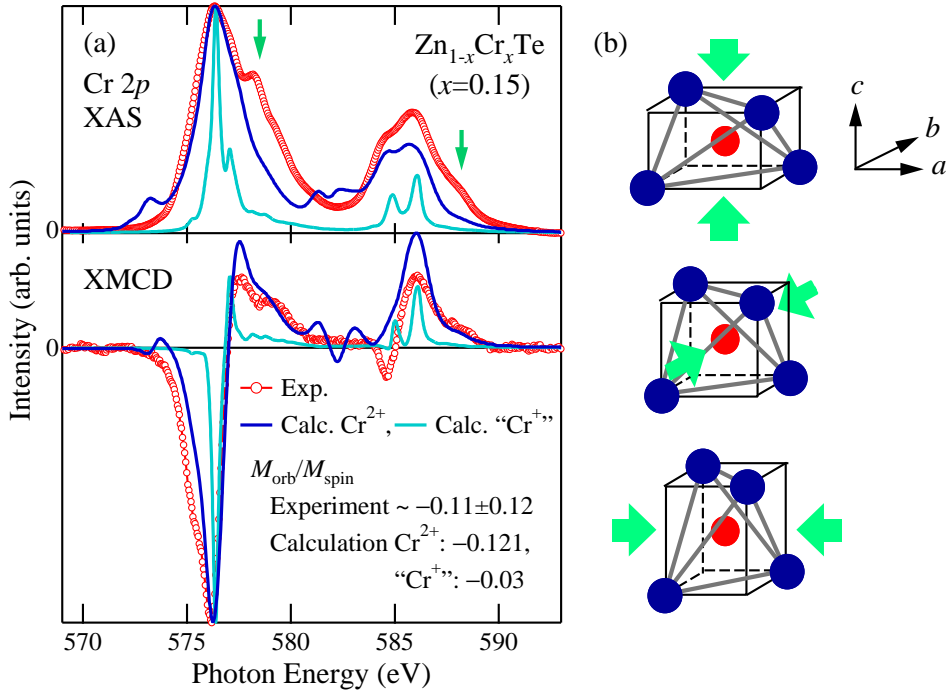}
\caption{Configuration-interaction cluster-model analysis for the Cr $2p$ XAS and XMCD spectra of Zn$_{1-x}$Cr$_x$Te. (a) Comparison between theory and experiment. Arrows denote the minority components. 
(b) Jahn-Teller distortions of the CrTe$_4$ tetrahedron. 
}
\label{ClusterCalc}
\end{center}
\end{figure}

Based on these considerations, we have performed CI cluster-model calculations on Cr$^{2+}$ in a $D_{2d}$ crystal field assuming the isotropic distribution of the Jahn-Teller axes and on Cr$^{2+}$ with negative $\Delta$. In the letter case, the ground state electronic configuration becomes $3d^5\underline{L}$ ($\underline{L}$ denotes a hole in the host valence band), which is referred to ``Cr$^{+}$" hereafter. Figure~\ref{ClusterCalc} shows comparison of the experimental spectra with calculated ones, where the calculated spectra are shifted to adapt the Cr $2p_{3/2}$ peak at 576 eV to the experimental one. The calculated spectra for ``Cr$^{+}$" are narrower than the experimental ones, where the electronic parameters are chosen as $\Delta=-1.0$, $U_{dd}=3.0$, and $(pd\sigma)=-0.75$ eV. Even if one chooses other sets of parameters with negative $\Delta$, the width of the calculated spectra does not change considerably. Furthermore, the value of $M_\mathrm{orb}/M_\mathrm{spin}$ of the calculated XMCD spectrum is too small compared to that of the experimental one. Therefore, the analysis suggest that the $d^{5}\underline{L}$ electronic configuration cannot explain the experimental spectra. 
On the other hand, the calculated XMCD spectrum for Cr$^{2+}$ agrees well with the experimental one, although there are some discrepancies between the Cr $2p$ XAS spectrum and the calculated one. The calculated $M_\mathrm{orb}/M_\mathrm{spin}=-0.121$ is similar to the $M_\mathrm{orb}/M_\mathrm{spin}$ value estimated from the XMCD sum rules $\sim -0.11\pm0.12$. 
The electronic structure parameters $\Delta$, $U_{dd}$, and ($pd\sigma$) are $4.0\pm0.5$, $3.5\pm0.5$, and $-1.1\pm0.1$ eV, respectively. 
Using these parameters, one can estimate the $p$-$d$ exchange constant $N\beta$, which is given by \cite{PRB_97_Mizokawa}
\begin{eqnarray}
N\beta &=& \frac{16}{S} \left( \frac{1}{\delta_{\mathrm{eff}}} - \frac{1}{\delta_{\mathrm{eff}}+4j} \right) 
  \left( \frac{1}{3}(pd\sigma) - \frac{2\sqrt{3}}{9} (pd\pi) \right)^2 \, \frac{1}{3} \nonumber \\
    &+& \frac{16}{S} \left( -\frac{1}{\delta_{\mathrm{eff}} + 6j} -  \frac{0.64}{-\delta_{\mathrm{eff}}+u'-j} \right)
       \left( \frac{1}{3}(pd\sigma) - \frac{2\sqrt{3}}{9} (pd\pi) \right)^2 \, \frac{2}{3}, 
\label{Nbeta}
\end{eqnarray}
where $u'$ and $j$ are Kanamori parameters, $\delta_{\mathrm{eff}} = \Delta_{\mathrm{eff}} + W_{V}/2$, $\Delta_\mathrm{eff}$ is the effective charge-transfer energy, and $W_{V}$ is the width of the host valence band. In the distorted CrTe$_4$ cluster, the first term of equation (\ref{Nbeta}) becomes dominant \cite{PRB_97_Mizokawa}. 
The $p$-$d$ exchange constant estimated from the parameters is $N\beta = 1.3\pm0.9$ eV ($>0$: ferromagnetic), consistent with the result of the MCD measurements \cite{Magneto-Optics}. 
The results indicate that the ferromagnetism in Zn$_{1-x}$Cr$_{x}$Te is caused by a single kind of Cr ions most likely substituting the Zn site. Other kinds of Cr ions are detected only in XAS and not in XMCD, since they are magnetically inactive at least in the present low magnetic fields $H \leqq 1$ T. These minority Cr ions may be antiferromagnetically coupled with each other. In fact, the line shape of the XMCD spectrum around the minority component peak position is slightly changed at high magnetic fields above $H \sim 2$ T \cite{APEX_08_Ishida}. 
The results indicate that the majority of Cr ions are divalent and are subjected to Jahn-Teller distortion with isotropic distribution of Jahn-Teller axes, giving support to the previous XMCD and PES results \cite{APEX_08_Ishida}.

\section{Resonant photoemission spectroscopy}
Although the previous measurements on Zn$_{1-x}$Cr$_{x}$Te ($x=0.045$) have shown a suppression of DOS near $E_\mathrm{F}$ \cite{APEX_08_Ishida}, it is not obvious whether the Zn$_{1-x}$Cr$_{x}$Te with high Cr concentration has a finite intensity at $E_\mathrm{F}$ or not. If double exchange interaction is dominant, the peak of the Cr $3d$ PDOS becomes broad with increasing Cr concentration. Therefore, the Cr-concentration dependence of the Cr $3d$ PDOS provides useful information about the origin of the ferromagnetism.

\begin{figure}[t!]
\begin{center}
\includegraphics[width=13.5cm]{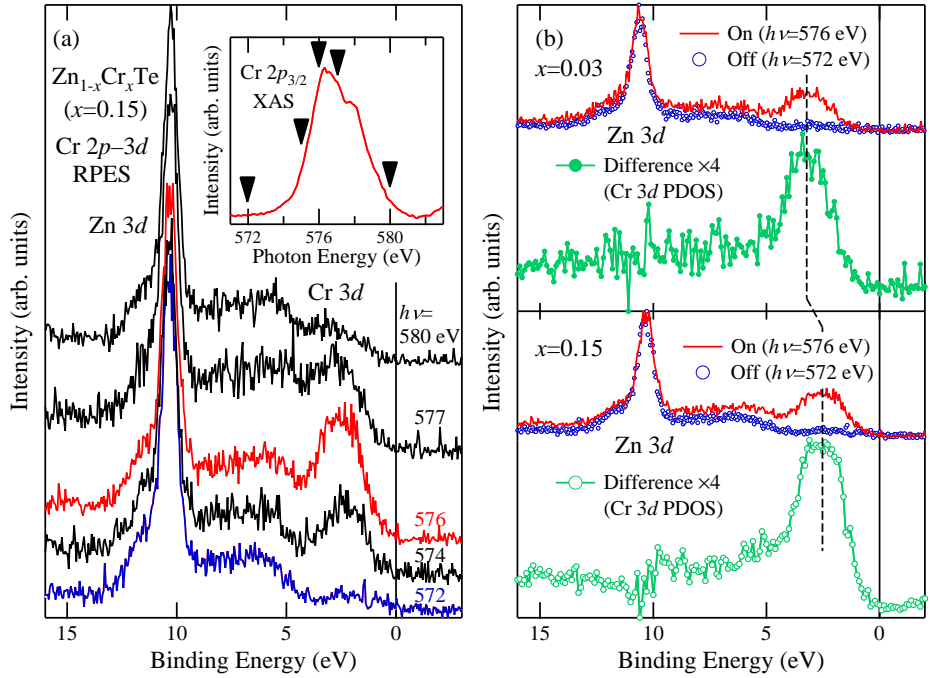}
\caption{Cr $2p \to 3d$ resonant photoemission spectra of Zn$_{1-x}$Cr$_x$Te.
(a) Cr $2p \to 3d$ resonant photoemission series. The photon energies ($h\nu$'s) are indicated on the Cr $2p_{3/2}$ XAS spectrum by triangles in the inset. 
(b) On- ($h\nu=576$ eV) and off-resonance ($h\nu=572$ eV) spectra of the $x=0.03$ and $x=0.15$ thin films. The difference spectra show the Cr $3d$ PDOS. 
}
\label{RPES}
\end{center}
\end{figure}

In order to obtain the Cr $3d$ PDOS in the valence-band region, RPES measurements have been performed using photons at the Cr $2p_{3/2}$ absorption edge. 
Since photoemission spectroscopy is a surface sensitive technique, the spectra involve signals from the Al capping layer, too. However, since the Cr ions are expected to exist in the Zn$_{1-x}$Cr$_x$Te layer and not in the Al capping layer, the spectral intensity enhanced at the Cr $2p$ edge is probably composed of signals from the Cr ions in the Zn$_{1-x}$Cr$_x$Te layer. 
We have observed a clear Cr $2p \to 3d$ resonance in the valence band as shown in Fig.~\ref{RPES}(a). Here, the on- and off-resonance spectra have been taken at $h\nu=576$ and 572 eV, respectively. 
The Cr $3d$ PDOS obtained by subtracting the off-resonance spectra from the on-resonance one shows a peak at the top of the valence band, indicating that if hole carriers are doped, the hole carriers should have $d$ character. 
The line shape of the PDOS is broad and almost independent of $x$, reflecting the strong hybridization between the host $s$,$p$ bands and the localized $d$ orbitals in Zn$_{1-x}$Cr$_{x}$Te. 
The $E_\mathrm{F}$ of the $x=0.15$ sample is located closer to the top of the valence band than that of the $x=0.03$ sample as shown in Fig.~\ref{RPES}(b). 
However, there is no density of states (DOS) at $E_\mathrm{F}$ for both the $x=0.03$ and 0.15 samples, corresponding to their insulating behaviors. Based on these findings, it is likely that the spectral suppression near $E_\mathrm{F}$ is originated from both the Jahn-Teller distortion and Coulomb interaction between Cr $3d$ electrons as proposed in the previous measurements \cite{APEX_08_Ishida}. 
Since the ferromagnetism realized by the double-exchange mechanism requires a finite DOS at $E_\mathrm{F}$ \cite{PR_51b_Zener, JJAP_01c_Sato}, the observations indicate that the double-exchange mechanism appears difficult in Zn$_{1-x}$Cr$_x$Te. The present results suggest that the ferromagnetic interaction between the spins of the $s$,$p$ band electrons and local moments of the $d$ orbitals is strong but the ferromagnetic interaction between the $d$ orbitals is shortranged because the top of the valence band has $d$ character \cite{JPCM_07_Belhadji}. In order to obtain further understanding of the ferromagnetic interaction, systematic XMCD measurements of ``carrier-doped'' (I and/or N doped) Zn$_{1-x}$Cr$_{x}$Te are highly desired.



\section{Conclusion}
We have performed XMCD and RPES measurements on the ferromagnetic DMS Zn$_{1-x}$Cr$_x$Te ($x=0.03$ and 0.15) thin films. 
The $x$ and $H$ dependences of the XMCD spectra provide experimental consistency with the magnetization measurements. The XMCD line shape is independent of $x$, $H$, $T$, and $\theta$, indicating that the Cr ions responsible for the ferromagnetism have a single chemical environment and a spatially isotropic electronic configuration. According to the XMCD sum-rule analysis and the $T$ dependence of XMCD, the ground states of the ferromagnetic Cr ions should involve the electronic configuration with the quenched orbital moment. The analysis using CI cluster-model calculation suggests that the Cr ions responsible for the ferromagnetism substitute for the Zn site and are subjected to Jahn-Teller distortion with isotropic distribution of the distortion axes, and there are two or more kinds of Cr ions. The minority Cr ions may be antiferromagnetically coupled with each other. The $p$-$d$ exchange constant $N\beta$ has been estimated from the electronic structure parameters and is consistent with MCD measurements. 
The Cr $3d$ PDOS at $E_\mathrm{F}$ is suppressed even at $x=0.15$, indicating that carrier-induced ferromagnetism appears difficult in Zn$_{1-x}$Cr$_{x}$Te. 
The suppression of spectral weight at $E_\mathrm{F}$ is probably caused by Jahn-Teller distortion and $d$-$d$ Coulomb interaction. 

From the report on the doping effects \cite{NatMater_07_Kuroda}, both the hole-carrier concentration and the inhomogeneous Cr distribution affect the ferromagnetic properties of Zn$_{1-x}$Cr$_{x}$Te thin film. In the present work, we have measured Zn$_{1-x}$Cr$_{x}$Te thin films having different Cr concentration. In this case, it is likely that the effects of the Cr inhomogeneity on the ferromagnetism are more dominant than these of the carrier concentration. Considering the fact that the ferromagnetic XMCD signal of Zn$_{1-x}$Cr$_x$Te is independent of Cr concentration and that Zn$_{1-x}$Cr$_x$Te is highly insulating, it is likely that the carrier-induced ferromagnetism is not important in Zn$_{1-x}$Cr$_{x}$Te and the inhomogeneous distribution of Cr atoms dominantly influences the ferromagnetic properties of Zn$_{1-x}$Cr$_x$Te. 


\section*{Acknowledgment}

We thank T. Mizokawa for fruitful discussions. This work was supported by a Grant-in-Aid for Scientific Research in Priority Area ``Creation and Control of Spin Current'' (19048012) from MEXT, Japan. The experiments at SPring-8 were approved by the Japan Synchrotron Radiation Research Institute (JASRI) Proposal Review Committee (Proposal No. 2006A3823). MK acknowledges support from the Japan Society for the Promotion of Science for Young Scientists.

\end{document}